%% file: main.tex
\documentclass[10pt, conference, letterpaper]{IEEEtran}

\usepackage{array, calc, color, multicol}
\usepackage{algorithm}
\usepackage[noend]{algorithmic} 
\usepackage{graphics, epstopdf, graphicx, epsfig, psfrag, epsf, subfigure}
\usepackage{amssymb, amsfonts, amsmath, times}
\usepackage{multicol,balance, verbatim}
\usepackage{textcomp, mathrsfs, stmaryrd, setspace}
\usepackage{amssymb}

\newcommand{\ie}{{\em i.e., }}
\newcommand{\Ie}{{\em I.e., }}
\newcommand{\eg}{{\em e.g., }}
\newcommand{\Eg}{{\em E.g., }}

\DeclareGraphicsExtensions{.eps, .pdf, .png, .jpg}   

\newcommand{\Sset}{\mathcal{S}}

\newcommand{\Nset}{\mathcal{N}}

\newcommand{\Lset}{\mathcal{L}}

\IEEEoverridecommandlockouts

\makeatletter
\newcommand{\oset}[2]{%
{\mathop{#2}\limits^{\vbox to -.5\ex@{\kern-\tw@\ex@
\hbox{\scriptsize #1}\vss}}}}
\makeatother

\begin{document}

\title{Device-Aware Routing and Scheduling in Multi-Hop Device-to-Device Networks}

\author{Yuxuan Xing and Hulya Seferoglu\\
{ ECE Department, University of Illinois at Chicago}\\
{ \tt yxing7@uic.edu, hulya@uic.edu}
}


\maketitle



\input{abstract}
\input{introduction}

\input{system}

\input{DARS}

\input{evaluation}
\input{related}
\input{conclusion}

\bibliographystyle{IEEEtran}
\bibliography{IEEEabrv,refs}

\end{document}

%% file: abstract.tex
\begin{abstract}
The dramatic increase in data and connectivity demand, in addition to heterogeneous device capabilities, poses a challenge for future wireless networks. One of the promising solutions is Device-to-Device (D2D) networking. D2D networking, advocating the idea of connecting two or more devices directly without traversing the core network, is promising to address the increasing data and connectivity demand. 
In this paper, we consider D2D networks, where devices with heterogeneous capabilities including computing power, energy limitations, and incentives participate in D2D activities heterogeneously. We develop (i) a {\em device-aware routing and scheduling algorithm} (DARS)  by taking into account device capabilities, and (ii) a multi-hop D2D testbed using Android-based smartphones and tablets by exploiting Wi-Fi Direct and legacy Wi-Fi connections.  We show that DARS significantly improves throughput in our testbed as compared to state-of-the-art. 
\end{abstract}

%



%% file: introduction.tex
\section{\label{sec:intro}Introduction}

The dramatic increase in the data and connectivity demand \cite{cisco_index, ericsson_report}, in addition to heterogeneous device capabilities, poses a challenge for future wireless networks. One of the promising solutions is Device-to-Device (D2D) networking. 

\begin{figure}[t!]
\centering
\vspace{-5pt}
\subfigure[The default operation]{ {\includegraphics[height=20mm]{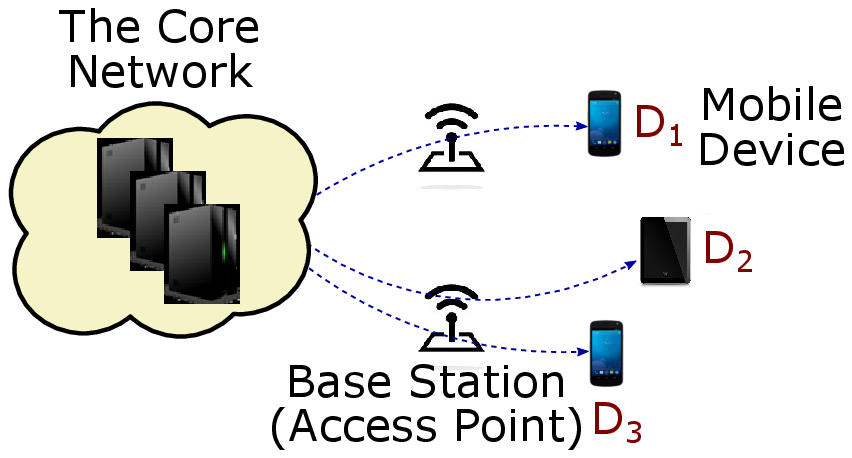}} } \hspace{2mm}
\subfigure[D2D connectivity]{ {\includegraphics[height=20mm]{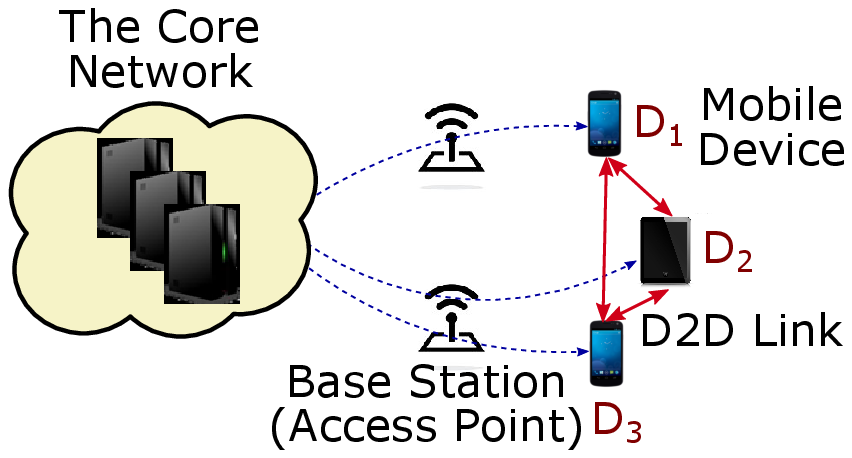}} } \hspace{2mm} \\
\subfigure[IoT get connectivity via D2D]{ {\includegraphics[height=23mm]{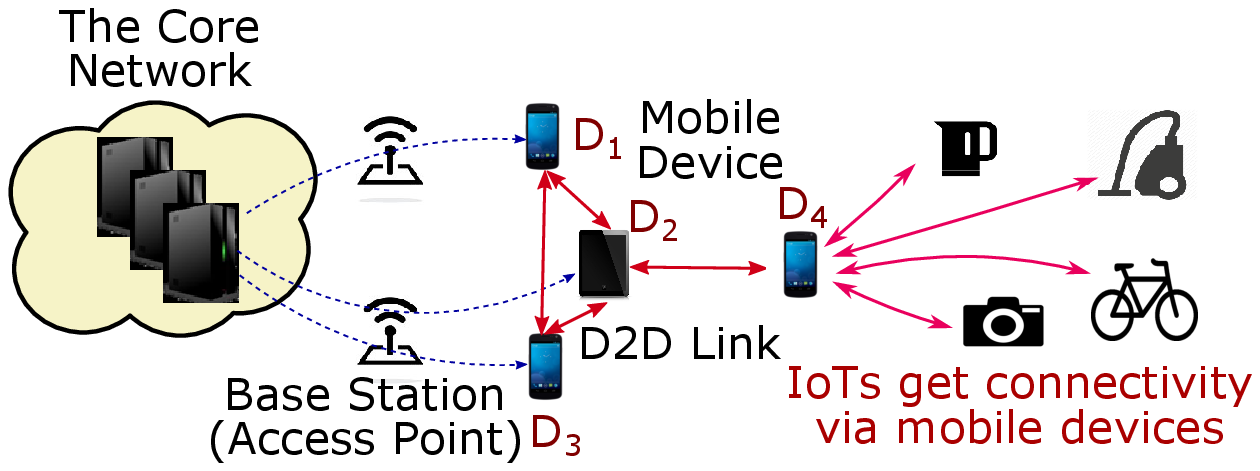}} }
\vspace{-5pt}
\caption{(a) The default operation for the Internet connection. (b) D2D connectivity: two or more mobile devices can be connected directly, \ie without traversing through the core network, if they are in close proximity by exploiting D2D connections such as Wi-Fi Direct or Bluetooth. (c) A number of devices (\eg IoT) seek connectivity via other devices (\eg mobile devices) using D2D connections. }
\vspace{-10pt}
\label{fig:intro_example}
\end{figure}

The default operation in current wireless networks is to connect each device to the Internet via its cellular or Wi-Fi interface, Fig.~\ref{fig:intro_example}(a). The D2D connectivity breaks this assumption: it advocates that two or more devices can be connected directly, \ie without traversing through an auxiliary device such as a base station if they are in close proximity \cite{survey_d2d}, Fig.~\ref{fig:intro_example}(b). D2D networks can be formed by exploiting D2D connections such as Wi-Fi Direct \cite{WiFiDirect} or Bluetooth. D2D networks are promising to address the ever-increasing number of devices as well as the demand for data and connectivity. 


Although D2D networking looks very promising to address the increasing data demand and the number of devices, and is expected to play a crucial role for the next generation networks, the following question is still open: How to design device-aware networking algorithms and protocols? 

In this paper, we consider a scenario where a number of devices, \eg mobile devices or Internet of Things (IoT), seek Internet connectivity via other devices using D2D connections as shown in Fig.~\ref{fig:intro_example}(c). In this context, it is possible to connect a device to the Internet via multiple hops, so it is crucial to determine which devices should forward packets, and how to make scheduling decisions. \Eg in Fig.~\ref{fig:intro_example}(c), there are two paths; $D_1-D_2-D_4$ and $D_3-D_2-D_4$, and it is crucial to determine the path that provides better connectivity. However, these decisions should be made by taking into account device capabilities. 

\begin{figure}[t!]
\centering
\subfigure[3-Node Line Topology]{ {\includegraphics[width=55mm]{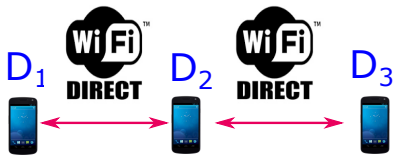}} } 
\subfigure[Rate vs Time]{ {\includegraphics[width=70mm]{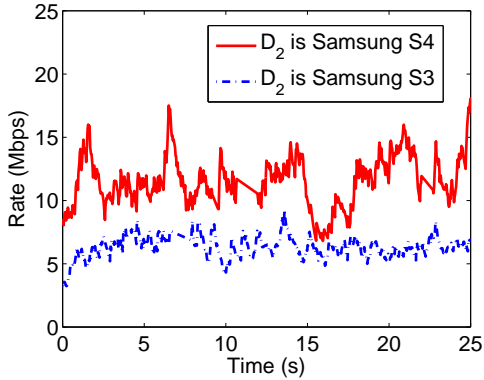}} } 
\vspace{-5pt}
\caption{{\bf Pilot study.} (a) The line topology, where $D_1$, $D_2$, and $D_3$ are connected via Wi-Fi Direct links. In our experiments, we used two Nexus 7 tablets, one Samsung S4 smartphone, and one Samsung S3 smartphone. All devices use Android as their operating systems, and Nexus 7 tablets are used as $D_1$ and $D_3$. (b) The rate between $D_1-D_3$ versus time when (i) $D_2$ is Samsung S4, and (ii) $D_2$ is Samsung S3. 
}
\vspace{-15pt}
\label{fig:motivating_example}
\end{figure}

\underline{{\em Pilot Study:}}
We developed a prototype for this pilot study as shown in Fig. \ref{fig:motivating_example}(a), where three devices $D_1$, $D_2$, and $D_3$ are connected as a line topology by exploiting the Wi-Fi Direct connections. In our pilot study, we used two Nexus 7 tablets, one Samsung S4 smartphone, and one Samsung S3 smartphone. 
Nexus 7 tablets are used as $D_1$ and $D_3$, and either Samsung S4 or Samsung S3 smartphone is used as $D_2$. 
In this setup, the capabilities of the intermediate device $D_2$, have direct impact on the transmission rate from $D_1$ to $D_3$. 
Our experimental results in Fig. \ref{fig:motivating_example}(b) show that when $D_2$ is Samsung S4 (a more powerful device as compared to Samsung S3), the transmission rate between $D_1-D_3$ is higher as compared to the case that $D_2$ is Samsung S3. As seen, the intermediate device with less computing power (Samsung S3) limits the transmission rate. 
\hfill $\Box$


Our pilot study shows that it is crucial to take into account device capabilities while designing D2D networking algorithms. Although our pilot study only focuses on the computing power, other parameters such as limited energy, human participation (or incentives), and bandwidth  should be taken into account. For example, it could be possible that the owner of $D_2$ may limit its participation in a D2D activity, which would eventually reduce the rate between $D_1-D_3$.

In this paper, we consider D2D networks, where devices with heterogeneous capabilities including computing power, energy limitations, and incentives participate in D2D activities heterogeneously. We first develop network utility maximization problem, and provide its solution. Then, based on the structure of the solution, we develop a {\em device-aware routing and scheduling algorithm} (DARS) that takes into account device capabilities. Furthermore, we design a multi-hop D2D testbed using Android-based smartphones and tablets by exploiting Wi-Fi Direct and legacy Wi-Fi connections. We evaluate DARS on this testbed. The following are the key contributions of this work: 

\begin{itemize}
  \item We consider a group of devices that form a multi-hop D2D network. We develop a network utility maximization (NUM) formulation of the device-aware framework, which provides a systematic approach to take into account device capabilities. We provide a decomposed solution of the NUM formulation, and based on the structure of the solution, we develop a stochastic {\em device-aware routing and scheduling algorithm} (DARS). 
  \item An integral part of our work is to understand the performance of DARS in practice. Towards this goal, we develop a testbed consisting of Nexus 5 smartphones, and Nexus 7 tablets. In this testbed, mobile devices can be configured in a multi-hop topology using Wi-Fi Direct interfaces. To the best of our knowledge, our implementation is the first that enables and supports real time multi-hop forwarding (instead of store and forward mechanism \cite{uscAndroidMultiHopImplementation} or using broadcast \cite{contentCentricRouting}) over Android-based mobile devices with Wi-Fi Direct. 
  \item We implemented DARS as well as the backpressure algorithm \cite{neelybook}, which is a state-of-the-art baseline on the testbed we developed. 
  The experimental results show that DARS brings significant performance benefits as compared to backpressure.
\end{itemize}

The structure of the rest of the paper is as follows. Section~\ref{sec:system} gives an overview of the system model and the problem formulation. Section~\ref{sec:DARS} presents DARS algorithm. Section~\ref{sec:implement} presents the implementation and evaluation of DARS. Section~\ref{sec:related} presents related work. Section~\ref{sec:conclusion} concludes the paper.

%% file: system.tex
\section{\label{sec:system}System Overview and Problem Formulation}
\subsection{\label{sec:system_det}System Overview}
We consider a multi-hop D2D network with mobile devices, where devices are connected to each other via D2D connections. In this setup, packets from a source device traverse potentially multiple devices before arriving to the destination device. Devices in this setup are capable of performing various tasks including routing, scheduling, and rate control. However, depending on device capabilities and configurations, the transmission rates vary. Our system model, and algorithm design capture this heterogeneity. In this section, we provide an overview of this setup and highlight some of its key characteristics.\footnote{We note that this section introduces our setup and assumptions needed for the theoretical development of our device-aware framework. We will revise some of these assumptions in Section~\ref{sec:implement} when we discuss implementation details of our algorithm in a testbed.}  

{\em Setup:} We consider a multi-hop D2D network, which consists of $N$ devices and $L$ edges, where $\Nset$ and $\Lset$ are the set of nodes and edges, respectively. We consider in our formulation and analysis that time is slotted, and $t$ refers to the beginning of slot $t$. 

{\em Sources and Flows:} Let $\Sset$ be the set of unicast flows between source and destination device pairs. Each flow $s \in \Sset$ generates  $A_s(t)$ packets at the application layer at time $t$. 

The packet arrivals are i.i.d. over the slots and the first and second moments of the arrival distribution is finite; \ie $\lambda_s = E[A_s(t)]$, and $E[A_s(t)^{2}]$. Packets are stored at the source device in an {\em initial buffer} in the application layer. Each flow $s$ is associated with rate $x_s$ and a utility function $g_{s}(x_{s})$, which we assume to be a strictly concave function of $x_{s}$ for our analysis purposes. Packets from the initial buffer are passed to the {\em main buffer} with rate  $x_{s}(t)$ at time $t$ and depending on the utility function $g_{s}(x_s(t))$. 

At time $t$, $f_{i,j}^{s}(t)$ packets from flow $s$ are passed from node $i$ to node $j$. The number of packets, \ie $f_{i,j}^{s}(t)$, are determined by device-aware framework by taking into account device capabilities.

\subsection{\label{sec:problem_formulation}Problem Formulation}
Now, we formulate our device-aware framework. Our objective is to determine $\boldsymbol x, \boldsymbol f$, where $\boldsymbol x = \{x_{s}\}_{s \in \Sset}$, and $\boldsymbol f = \{f_{i,j}^s\}_{s \in \Sset, (i,j) \in \Lset}$, by maximizing the total utility function; $\sum_{s \in \Sset} g_{s}(x_s)$ subject to the constraints\footnote{Note that, in this section, we optimize the average values of the parameters defined in Section~\ref{sec:system_det}. Thus, by abuse of notation, we use a variable, \eg $\phi$ as the average value of $\phi(t)$  if both $\phi$ and $\phi(t)$ refers to the same parameter.} 
\begin{align} \label{opt:eq1}
& \sum_{j \in \Nset} f_{i,j}^{s} - \sum_{j \in \Nset} f_{j,i}^{s} = x_s1_{[i=o(s)]}, \mbox{ } \forall s \in \Sset, i \in \Nset \nonumber \\
& \sum_{s \in \Sset} \sum_{i \in \Nset} f_{i,j}^{s} \leq \min \{ R_{P}^j, R_{E}^j, R_{W}^j\}, \forall j \in \Nset  \nonumber \\
& \boldsymbol f \in \boldsymbol \Gamma
\end{align}

The first constraint in  (\ref{opt:eq1}) is the flow conservation at device $i$ and for flow $s$, where $\sum_{j \in \Nset} f_{j,i}^{s} + x_s1_{[i=o(s)]}$ is the arrival rate of flow $s$ to node $i$, while $\sum_{j \in \Nset} f_{i,j}^{s}$ is the departure rate.

The second constraint captures device capabilities. The arrival rate to device $j$, \ie $\sum_{s \in \Sset} \sum_{i \in \Nset} f_{i,j}^{s}$ should be supported by the device, where $R_{P}^j$ is the maximum rate that device $j$ can support (receive and transmit) with its computing power, while $R_{E}^j$ and $R_{W}^j$ are the rates that device $j$ can support by with its energy and incentives, respectively.

The last constraint in  (\ref{opt:eq1}) is the feasibility constraint, where $\boldsymbol \Gamma$ is the set of all feasible rates that can be in the network. Thus, $\boldsymbol f$ should be an element of $\boldsymbol \Gamma$. 

Although the solution of (\ref{opt:eq1}) provides a device-aware routing and scheduling, the solution is not practical, because it requires an active involvement of all devices in D2D network even if a device does not prefer any involvement. For example, even if a node $j$ has very small $\min \{ R_{P}^j, R_{E}^j, R_{W}^j\}$, it needs to periodically update the other devices in the network about its status (whether it can relay packets or not), which is not practical and introduces overhead. Thus, we modify the problem in (\ref{opt:eq1}) so that the solution can be more practical. 

Our first step is to explicitly involve link rates in the formulation. Assume that  $R_{i,j}$ is the transmission rate between nodes $i,j$ and $\tau_{i,j}$ is the percentage of time that the link $i-j$ is used. Then, we can express $ f_{i,j}^{s} = R_{i,j} \tau_{i,j}^s$, $\forall i \in \Nset, j \in \Nset, s \in \Sset$. This translates the constraints in (\ref{opt:eq1}) to $\sum_{j \in \Nset} R_{i,j}\tau_{i,j}^{s} - \sum_{j \in \Nset} R_{j,i} \tau_{j,i}^{s} = x_s1_{[i=o(s)]}, \mbox{ } \forall s \in \Sset, i \in \Nset$, $\sum_{s \in \Sset} \sum_{i \in \Nset} R_{i,j}\tau_{i,j}^{s} \leq \min \{ R_{P}^j, R_{E}^j, R_{W}^j\}, \forall j \in \Nset$ , and $\boldsymbol \tau \in   \boldsymbol  \Gamma_{\tau}$, where $\boldsymbol \tau = \{\tau_{i,j}^s\}_{s \in \Sset, (i,j) \in \Lset}$, and $\boldsymbol \Gamma_{\tau}$ is the set of all feasible link schedules, so $\boldsymbol \tau \in   \boldsymbol  \Gamma_{\tau}$ should hold. 

The next step is to create a new variable $\gamma_{i,j}^{s}$ as $\gamma_{i,j}^{s}$ $=$ $\tau_{i,j}^{s}$ $\frac{R_{i,j}}{\min \{ R_{E}^j, R_{W}^j, R_{P}^j\}}$ assuming that $\min \{ R_{P}^j, R_{E}^j, R_{W}^j\}$ is positive, at least slightly larger than 0. Then, the constraints become
\begin{align} \label{opt:eq3}
& \sum_{j \in \Nset} \gamma_{i,j}^{s} \min \{ R_{P}^j, R_{E}^j, R_{W}^j\} - \sum_{j \in \Nset}  \gamma_{j,i}^{s} \min \{ R_{E}^i, R_{W}^i, R_{P}^i\} \nonumber \\
& = x_s1_{[i=o(s)]}, \mbox{ } \forall s \in \Sset, i \in \Nset \nonumber \\
& \sum_{s \in \Sset} \sum_{i \in \Nset} \gamma_{i,j}^{s} \leq 1, \forall j \in \Nset   \mbox{   and    } \boldsymbol \gamma \in   \boldsymbol  \Gamma_{\gamma},
\end{align} where $\boldsymbol \gamma = \{\gamma_{i,j}^s\}_{s \in \Sset, (i,j) \in \Lset}$, and $\boldsymbol  \Gamma_{\gamma}$ is the set of feasible $\boldsymbol \gamma$'s. Next, we provide a solution to the problem of maximizing the total utility; $\sum_{s \in \Sset} g_{s}(x_s)$ subject to the constraints in (\ref{opt:eq3}). 


\subsection{\label{sec:NUM_solution}Solution}
Lagrangian relaxation of the first constraint of (\ref{opt:eq3}) gives the following Lagrange function: 
\begin{align} \label{relax:eq1}
L = & \sum_{s \in \Sset} g_s(x_s) - \sum_{s \in \Sset} \sum_{i \in \Nset} u_i^s ( \sum_{j \in \Nset} \gamma_{i,j}^{s} \min \{ R_{E}^j, R_{W}^j, R_{P}^j\} - \nonumber \\
& \sum_{j \in \Nset} \gamma_{j,i}^{s} \min \{ R_{E}^i, R_{W}^i, R_{P}^i\} -  x_s 1_{[i=o(s)]} )
\end{align} where $u_i^s$ is the Lagrange multiplier. 
The Lagrange function is expressed as 
\begin{align} \label{relax:eq1}
L = & \sum_{s \in \Sset} [g_s(x_s) - u_{o(s)}^{s} x_s] + \sum_{s \in \Sset} \sum_{i \in \Nset} \sum_{j \in \Nset} \gamma_{i,j}^{s} \min \{ R_{E}^j, R_{W}^j, \nonumber \\
& R_{P}^j\} (u_i^s - u_j^s). 
\end{align}
This Lagrange function is decomposed into two sub-problems: (i) rate control, and (ii) routing and scheduling. If we solve the Lagrangian function with respect to $x_s$, we have an optimization problem: $\max_{\boldsymbol x} \sum_{s \in \Sset} [g_s(x_s) - u_{o(s)}^{s} x_s] $, which is the rate control part. On the other hand, the routing and scheduling part solves the following optimization problem
\begin{align} \label{eq:routing3}
\max_{\boldsymbol \gamma} \mbox{ } & \sum_{s \in \Sset} \sum_{i \in \Nset} \sum_{j \in \Nset} \gamma_{i,j}^{s} \min \{ R_{E}^j, R_{W}^j,  R_{P}^j\}  (u_i^s - u_j^s) \nonumber \\
\mbox{s.t.} \mbox{ }  &  \sum_{s \in \Sset} \sum_{i \in \Nset} \gamma_{i,j}^{s} \leq 1, \forall j \in \Nset  \mbox{   and    }  \boldsymbol \gamma \in {\boldsymbol \Gamma_\gamma}.
\end{align} 
Note that the solution of (\ref{eq:routing3}) is easier as compared to the solution of (\ref{opt:eq1}), because it reduces to selecting the link $i-j$, which maximizes $ \min \{ R_{E}^j, R_{W}^j,  R_{P}^j\}  (u_i^s - u_j^s)$ among all feasible schedules of links. Based on this idea, we will develop our device-aware stochastic routing and scheduling algorithm in the next section.

%% file: DARS.tex
\section{\label{sec:DARS} DARS: Device-Aware Routing and Scheduling}
Now, we design DARS, which has (i) rate control, (ii) routing and scheduling, and (iii) queue evolution parts, based on the solutions developed in Section~\ref{sec:NUM_solution}.


\underline{Device-Aware Routing and Scheduling Algorithm (DARS):}
\begin{itemize}
 \item {\em Rate Control:} At slot $t$, the rate controller at node $o(s)$ determines the number of packets that should be passed from the initial buffer to the main buffer according to
\begin{align} \label{eq:rate_control}
\max_{\boldsymbol {{x}}} & \sum_{s \in \Sset} [Mg_s(x_s(t)) - U_{o(s)}^{s} x_s(t)] \nonumber \\
\mbox{s.t. } &  x_s(t) \leq R_{\max},
\end{align} where $U_{o(s)}^{s}$ is the queue that stores packets from flow $s$ at node $o(s)$, $R_{\max}$ is a positive constant larger than the transmission rate from device $o(s)$, and $M$ is a large positive constant. Note that flow control algorithm in (\ref{eq:rate_control}) is designed based on the structure of the rate control solution in Section~\ref{sec:NUM_solution}.


\item {\em Routing and Scheduling:}  At slot $t$, device $j$ determines the number of packets that it can receive, process, and forward according to
\begin{align} \label{eq:routing}
\max_{\boldsymbol {{\gamma}}} & \sum_{s \in \Sset} \sum_{i \in \Nset} \gamma_{i,j}^{s}(t) \min \{ R_{P}^j, R_{E}^j, R_{W}^j\} [ U_{i}^{s}(t) - U_{j}^{s}(t)] \nonumber \\
\mbox{s.t. } & \sum_{s \in \Sset} \sum_{i \in \Nset} \gamma_{i,j}^s(t) \leq 1, \mbox{   and    } {\boldsymbol \gamma(t)} \in {\boldsymbol \Gamma_\gamma (t)}
\end{align} where $U_{i}^{s}(t)$ and $U_{j}^{s}(t)$ are queue sizes at nodes $i$ and $j$, respectively. After the value of $\gamma_{i,j}^{s}(t)$ is determined, if $\gamma_{i,j}^{s}(t) = 1$, $f_{i,j}^{s}(t)$ is set to $f_{i,j}^{s}(t) = F_{\max}$, where $F_{\max}$ is a positive constant larger than the transmission rate from device $i$ to device $j$, as well as larger than $\min \{ R_{P}^j, R_{E}^j, R_{W}^j\}$. Otherwise, \ie if $\gamma_{i,j}^{s}(t) = 0$, then $f_{i,j}^{s}(t) = 0$. The solution in (\ref{eq:routing}) has two strengths: (i) it takes into account device capabilities, \ie $\min \{ R_{P}^j, R_{E}^j, R_{W}^j\}$, and  (ii) each device $j$ makes its own decision on how much data it can handle (route \& schedule), which is fundamentally different than the classical backpressure \cite{tass1, neelymoli, neelybook}, where each device $j$ should do its best to route and schedule any amount of data it receives. Also, note that the solution of (\ref{eq:routing}) is both a routing decision as it determines the next hops, and a scheduling decision as it determines which links to activate (the ones $\gamma_{i,j}^{s}(t) = 1$ are activated).

\item {\em Queue Evolution:} The evolution of the queue $U_{i}^s(t)$ at time $t$ is as follows;
\begin{align} \label{eq:U_evol}
& U_i^s(t+1) \leq \max [U_i^s(t) - \sum_{j \in \Nset} f_{i,j}^{s}(t), 0] + \sum_{j \in \Nset} f_{j,i}^{s}(t) \nonumber \\
& + x_s(t)1_{[i=o(s)]}
\end{align} where $o(s)$ is the source node of flow $s$ and $1_{[i=o(s)]}$ is an indicator function, which is $1$ if $i=o(s)$, and $0$, otherwise. Note that (\ref{eq:U_evol}) is an inequality, because the actual amount of data arriving to the queue may be smaller than $\sum_{j \in \Nset} f_{j,i}^{s}(t) + x_s(t)1_{[i=o(s)]} $.
\end{itemize}

In Section~\ref{sec:implement}, we will provide the implementation details and evaluation of DARS in a real testbed. Yet, before delving into that, we present simulation results of DARS as compared to backpressure \cite{neelymoli} in an idealized setup.

\begin{figure}[t!]
\centering
{ {\includegraphics[height=30mm]{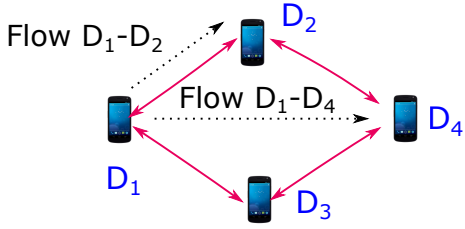}} }
\caption{Diamond topology for simulations.
}
\vspace{-10pt}
\label{fig:diamond_topology}
\end{figure}

\begin{figure}[t!]
\centering
\subfigure[Rate vs. $\min$ $\{R_P^2,$ $R_E^2,$ $R_W^2\}$]{ {\includegraphics[height=40mm]{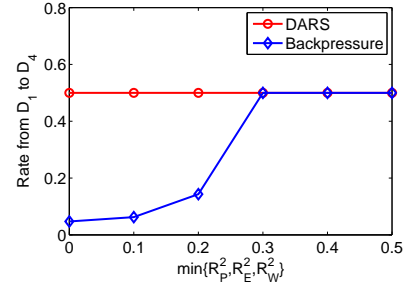}} }
\subfigure[Rate vs. loss probability]{ {\includegraphics[height=40mm]{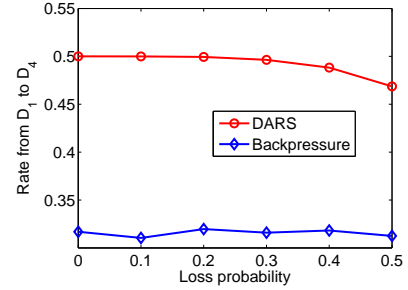}} }
\caption{Diamond topology with one flow from $D_1$ to $D_4$. (a) Rate of flow $D_1 - D_4$ versus $\min$ $\{$ $R_P^2,$ $R_E^2,$ $R_W^2\}$, where links are not lossy. (b) Rate of flow $D_1 - D_4$ versus loss probability, where all links are lossy. In both simulations, $M=200$, $R_{\max} = 1$, $F_{\max} = 1$.
}
\vspace{-10pt}
\label{fig:matlab_sims_one_flow}
\end{figure}

We first consider a diamond topology shown in Fig.~\ref{fig:diamond_topology}, where there is a flow from $D_1$ to $D_4$. For this scenario, Fig.~\ref{fig:matlab_sims_one_flow}(a) shows the flow rate versus $\min$ $\{R_P^2,$ $R_E^2,$ $R_W^2\}$, when $\min$ $\{R_P^i,$ $R_E^i,$ $R_W^i\}=1$, for $i=1,3,4$, and links are not lossy. Fig.~\ref{fig:matlab_sims_one_flow}(b) shows the flow rate versus loss probability (all links are lossy), when $\min$ $\{R_P^2,$ $R_E^2,$ $R_W^2\}=0.1$ and $\min$ $\{R_P^i,$ $R_E^i,$ $R_W^i\}=1$, for $i=1,3,4$. In both simulations, $M=200$, $R_{\max} = 1$, $F_{\max} = 1$. The results show that DARS significantly improves over backpressure (implemented according to \cite{neelymoli}), because DARS takes into account device capabilities while backpressure does not. \Ie even if packets and links are scheduled by backpressure, they may not be realized due to the device ($D_2$ in this simulation) bottleneck.

\begin{figure}[t!]
\centering
\subfigure[Rate vs. $\min$ $\{R_P^2,$ $R_E^2,$ $R_W^2\}$]{ {\includegraphics[height=40mm]{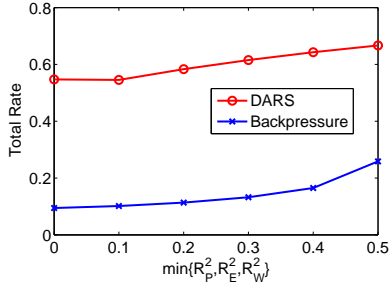}} }
\subfigure[Rate vs. loss probability]{ {\includegraphics[height=40mm]{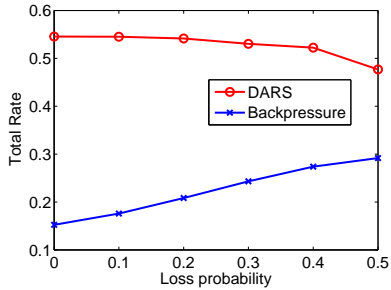}} }
\caption{Diamond topology with two flows; one from $D_1$ to $D_4$, and another from $D_1$ to $D_2$. (a) Total rate (of both flows) versus $\min$ $\{R_P^2,$ $R_E^2,$ $R_W^2\}$. (b) Total rate (of both flows) versus loss probability, where all links are lossy. In both simulations, $M=200$, $R_{\max} = 1$, $F_{\max} = 1$.
}
\vspace{-10pt}
\label{fig:matlab_sims_two_flow}
\end{figure}

Next, we consider the diamond topology shown in Fig.~\ref{fig:diamond_topology} for two flows; one from $D_1$ to $D_2$, and another from $D_1$ to $D_4$. Considering the same parameters of the one-flow scenario above, we have results as shown in Fig.~\ref{fig:matlab_sims_two_flow}. As seen, DARS dramatically improves over backpressure as in the one-flow case thanks to taking into account device capabilities.

%% file: evaluation.tex
\section{\label{sec:implement} Implementation Details and Evaluation}

In this section, we present the implementation details of our testbed. First, we start with how we create a multi-hop topology with Android-based devices using Wi-Fi Direct. Then, we present our DARS implementation on our testbed.

\subsection{\label{sec:multi_hop} Creating Multi-Hop Topology with Android Devices}

\begin{figure*}[t!]
\centering
\subfigure[Star topology with Wi-Fi Direct]{ {\includegraphics[height=24mm]{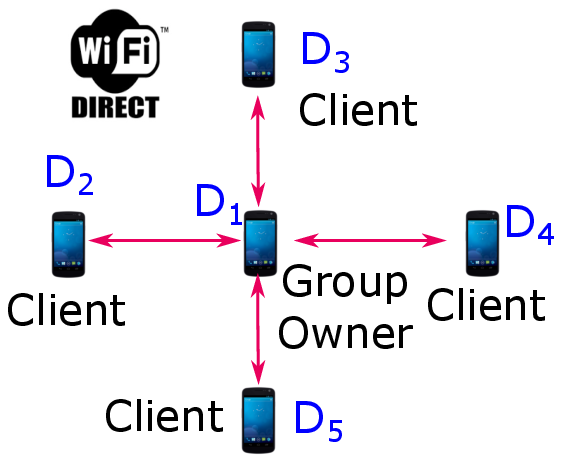}} } \hspace{5pt}
\subfigure[Line topology in our testbed]{ {\includegraphics[height=16mm]{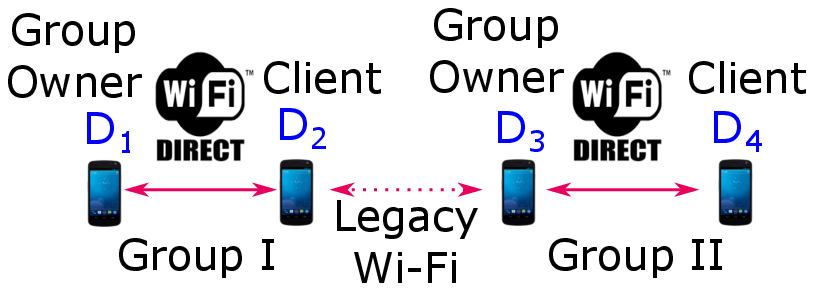}} } \hspace{5pt}
\subfigure[Diamond topology in our testbed]{ {\includegraphics[height=26mm]{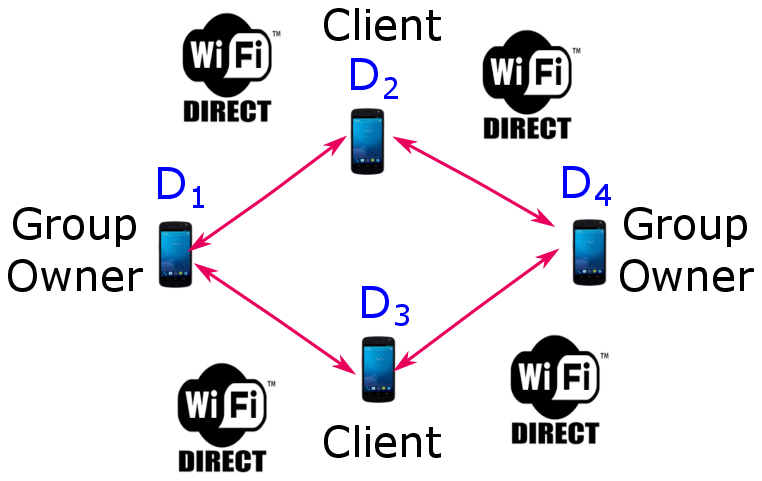}} }
\vspace{-5pt}
\caption{(a) Diamond topology for simulations. (a) Star topology. A device that receives a connection request first becomes a group owner and center of the star topology, \ie $D_1$, and all other devices, \ie $D_2$ to $D_5$, connect to the group owner. In this setup, there can be only one group owner, and a device cannot act as both group owner and a client simultaneously. (b) Line topology with two groups. Group I and Group II are created using Wi-Fi Direct. $D_1$ and $D_3$ are group owners, and  $D_2$ and $D_4$ are clients of Group I and Group II, respectively. $D_2$ is connected to $D_3$ using legacy Wi-Fi interface. This creates a 4-node line topology. (c) Diamond topology. $D_1$ and $D_4$ are group owners, and $D_2$ and $D_3$ are clients in Wi-Fi Direct groups. $D_2$ and $D_3$ are connected to $D_1$ via Wi-Fi Direct interface, and connect to $D_4$ using their legacy Wi-Fi interfaces.
}
\vspace{-10pt}
\label{fig:topologies}
\end{figure*}

{\em Wi-Fi Direct:} Our approach to create multi-hop topology using Android-based devices is to employ Wi-Fi Direct connections \cite{WiFiDirect}. However, existing Wi-Fi Direct implementation in Android-based devices only supports a star topology as shown in Fig.~\ref{fig:topologies}(a), but not any other multi-hop topology.

{\em Our Approach:} We use the star topology of Wi-Fi Direct shown in Fig.~\ref{fig:topologies}(a) as our basic constructing unit for creating multi-hop topologies. In particular, multiple groups are constructed using the star topology (\ie each group is a star topology), and these groups are connected to each other. Connecting multiple groups (star topologies) is quite challenging, because the star topology of Wi-Fi Direct is constructed in a way that one device (center of the topology) is a group owner (GO), and the other devices are clients. In this setup, a device cannot act as both a group owner and a client simultaneously, which makes connecting multiple groups to each other prohibitively difficult.

In our testbed, we use legacy Wi-Fi interface to connect multiple groups. Let us consider Fig.~\ref{fig:topologies}(b), where there are two groups; Group I: $D_1$, $D_2$ and Group II: $D_3$, $D_4$. In this example, $D_1$ and $D_3$ are group owners, and  $D_2$ and $D_4$ are clients of Group I and Group II, respectively. Let us assume that our goal is to connect $D_2$ and $D_3$. In existing Wi-Fi Direct, $D_2$ cannot connect to $D_3$ as $D_2$ can only connect to its group owner (which is $D_1$) via Wi-Fi Direct interface. On the other hand, $D_3$ cannot connect to $D_2$ as a client, because $D_3$ is already a group owner of Group II, so it cannot be a client of $D_2$. Therefore, our approach is to use legacy Wi-Fi interface of $D_2$ to connect to $D_3$. This connection is possible as $D_2$ will see $D_3$ as an access point of the legacy Wi-Fi connection. Thus, $D_2$ can connect to $D_3$, which provides a line topology consisting of 4 nodes, which was not possible by using only existing Wi-Fi Direct setup.

Similarly, we can create other multi-hop topologies. For example, we can create a diamond topology as shown in Fig.~\ref{fig:topologies}(c), where $D_1$ and $D_4$ are group owners, and $D_2$ and $D_3$ are clients in Wi-Fi Direct groups. $D_2$ and $D_3$ are connected to $D_1$ via Wi-Fi Direct interface, and connect to $D_4$ using their legacy Wi-Fi interfaces.

Our approach of using legacy Wi-Fi interfaces simultaneously with Wi-Fi Direct interfaces is challenging, because both legacy Wi-Fi interface and Wi-Fi Direct interfaces are actually using the same means of communication interface in the MAC layer, which is 802.11. Thus, if we naively open both legacy Wi-Fi and Wi-Fi Direct interfaces, only one of them will operate due to IP addressing conflicts. For example, $D_2$ would transmit data to $D_1$ even if it means to transmit to $D_3$ in Fig.~\ref{fig:topologies}(b). We provide a solution to this problem in a simple way (\ie without rooting mobile devices). In particular, we use a class called ConnectivityManager in Android API, which provides instances (objects of the class) of all active network interfaces on each device. Thus, we access the instance of legacy Wi-Fi interface, and bind it with transmission sockets TCP or UDP. This approach eliminates addressing issues and conflicts between legacy Wi-Fi and Wi-Fi Direct interfaces.

\subsection{DARS Implementation}
In this section, we present how DARS is implemented over our multi-hop testbed described in Section~\ref{sec:multi_hop}.

{\em Devices:} We implemented a testbed of the different topologies including line topology, diamond topology using real mobile devices, specifically Android 5.1.1 based Nexus 5 smartphones and Nexus 7 tablets.

\begin{figure}
\centering
\scalebox{.45}{\includegraphics{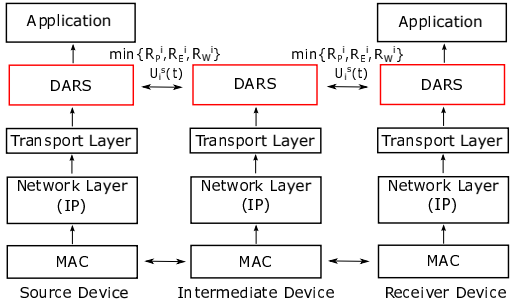}}
\vspace{-5pt}
\caption{DARS operations at end-points and intermediate nodes.}
\label{fig:protocolStack}
\vspace{-15pt}
\end{figure}

{\em Integration to the Protocol Stack:} We implemented DARS as a slim layer between transport and application layers as demonstrated in Fig.~\ref{fig:protocolStack}. In other words, we implemented DARS on top of TCP. This kind of implementation has benefits as (i) mobile devices do not require rooting, and (ii) our DARS codes could be easily transferred to mobile devices using other operating systems such as iOS.

{\em Virtual Slots:} As mentioned in Section~\ref{sec:system}, DARS uses slots to make transmission decisions. By following the theory, in our implementation, we divided the time into virtual slots. Each decision is made at the start of the slot. We set slot durations to 50msec.

{\em Multiple Threads:} Three sets of threads operate at each device simultaneously to perform the tasks of rate control, routing and scheduling, and actual data transmission.

The first set of threads are implemented for the rate control, so we call them rate control threads. In particular, the rate control thread at the source device $o(s)$ reads data bytes from a file, packetizes them and inserts them into the transmission queue $U_{o(s)}^s$. The rate of reading packets from the file is determined according to the rate control algorithm in (\ref{eq:rate_control}). Note that if a device is the source of two flows, a rate control thread is created for each flow.

The second set of threads determine how many packets should be transmitted from $U_{i}^s$ to other devices. Thus, these threads are called routing and scheduling threads. This part implements (\ref{eq:routing}). For example, in the diamond topology in Fig.~\ref{fig:topologies}(d), at each slot, $D_1$ determines whether it should transmit packets to $D_2$, or $D_3$, or none of them.

The final set of threads make actual packet transmissions possible, so we call them transmission threads. A transmission thread is constructed for each neighboring node. For example, in the diamond topology Fig.~\ref{fig:topologies}(c), $D_1$ constructs two transmission threads for $D_2$ and $D_3$. Packets, whose number is determined by the routing and scheduling thread, are received by this thread and inserted into queues that we call socket queues. The transmission threads will dequeue packets from the socket queues and pass them to TCP sockets. Another task of transmission threads is to receive queue size information from neighboring nodes.

{\em Information Exchange:} Our implementation is lightweight in the sense that it limits control information exchange among mobile devices. The control information that is transmitted by node $i$ is $U_{i}^{s}(t)$ and $\min \{ R_{P}^i, R_{E}^i, R_{W}^i\}$, and this information is transmitted to only $i$'s neighbors. These control packets are transmitted periodically at every 50msec.

{\em Calculating $\min \{ R_{P}^i, R_{E}^i, R_{W}^i\}$:} Each node $i$, based on its computing power, energy level, and incentives (willingness), calculates its rate. For example, if node $i$ has limited energy, then it limits $R_{E}^i$ to 1Mbps even if it can support up to 20Mbps. Thus, in our implementation, every device calculates its own rate, and exchange this information with its neighbors.

{\em Test Environment:} We conducted our experiments in a lab environment where several other Wi-Fi networks were operating in the background. We located all the devices at varying distances, and we have evaluated device-centric routing and scheduling. Next, we present our evaluation results.

\subsection{Evaluation Results}
\begin{figure}[t!]
\centering
\vspace{-5pt}
\subfigure[Line topology results]{ {\includegraphics[height=40mm]{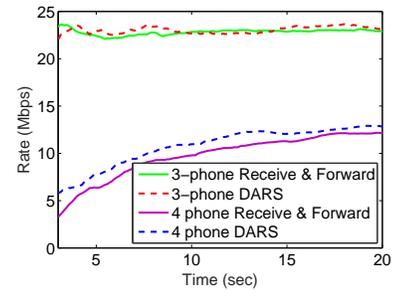}} } 
\subfigure[Diamond topology results]{ {\includegraphics[height=40mm]{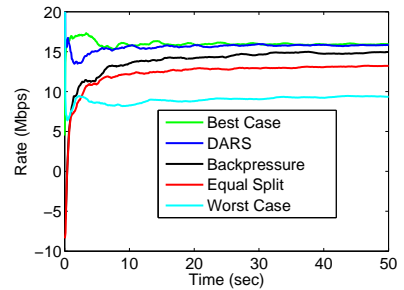}} }
\vspace{-5pt}
\caption{
(a)  Line topology (as shown in Fig.~\ref{fig:topologies}(c)) results with three and four devices. (b) Diamond topology (Fig.~\ref{fig:topologies}(d)) results. }
\vspace{-20pt}
\label{fig:results}
\end{figure}

Now, we focus on evaluating the performance of DARS. Fig.~\ref{fig:results}(a) shows data rate versus time graph for three-node and four-node line topologies shown in Fig.~\ref{fig:topologies}(b), where a a flow is transmitted from $D_1$ to $D_4$. The receive \& forward algorithm is the baseline in this scenario, where the intermediate nodes just receive packets and forward. As seen, the performance of DARS is close to receive \& forward. Note that since there is no routing and scheduling diversity in the line topology, the receive \& forward provides the best performance. The results prove that DARS does not introduce too much overhead into the system.  

Fig.~\ref{fig:results}(b) shows the transmission rate for the diamond topology shown in Fig.~\ref{fig:topologies}(c), where a flow is transmitted from $D_1$ to $D_4$. In this setup, $D_1$, $D_2$, and $D_4$ are Nexus 5 smartphones, and $D_3$ is a Nexus 7 tablet. The computing power of Nexus 5 smartphones is better than Nexus 7 tablet. In particular, Nexus 5 supports two times faster rate as compared to Nexus 7. Thus, DARS should prefer $D_1 - D_2 - D_4$ route instead of $D_1 - D_3 - D_4$ when $D_1$ is the source and $D_4$ is the receiver. The simulation results support this as further explained next.

The best case scenario in Fig.~\ref{fig:results}(b) is the case that $D_1 - D_2 - D_4$ route is selected a priori. As seen, DARS performs very close to the best case scenario, which shows the efficiency of our algorithm. Backpressure is the implementation of the scheme proposed in \cite{neelymoli}. As seen, DARS improves over backpressure (up to \%15), because it takes into account device specific properties, while backpressure does not. Equal Split is another baseline, which allows transmitting data over both paths in the diamond topology. As long as TCP supports transmissions, packets are simultaneously transmitted over both links. As seen, DARS significantly improves as compared to this baseline. Finally, the worst case scenario is the case that $D_1 - D_3 - D_4$ route is selected a priori, which is included in the results for completeness.

%% file: related.tex
\section{\label{sec:related} Related Work}
The idea of exploiting D2D connectivity is very promising to improve throughput and reduce delay, so it has found several applications in the literature. For example, opportunistic D2D connections is often used for the purpose of (i) offloading cellular networks \cite{micro1}, \cite{micro2}, \cite{micro3}, (ii) content dissemination among mobile devices \cite{micro4,micro5}, and (iii) cooperative video streaming over mobile devices \cite{micro6, micro7}. As compared to this line of work, we focus on developing {\em device-aware routing and scheduling} algorithm over multi-hop D2D networks by taking into account device capabilities such as  computing power, energy, and incentives.

Our approach in this work involves using network utility maximization to characterize the system as it is promising to understand how different layers and/or algorithms, such as flow control, routing, and scheduling should be designed and optimized \cite{tutorial_doyle}, \cite{tutorial_lin}. However, we formulate the NUM framework considering device capabilities to develop device-aware framework. Second, we develop DARS. In that sense, our approach is similar to the line of work emerged after the pioneering work in \cite{tass1}, \cite{tass2}, \cite{neelymoli}. However, our focus is on incorporating device capabilities in the framework. Furthermore, we develop a testbed of our algorithm using real devices, which was not the focus of the previous work.

Multi-hop data transmission testbed using Wi-Fi Direct over mobile devices has been considered in \cite{uscAndroidMultiHopImplementation}. In this work, intermediate devices receive a whole file first, and then transmits it to other devices. As compared to this work, our implementation makes simultaneous transmission and reception possible, so there is no need to wait to receive a complete file before starting to transmit it to a next hop. More similar work to ours is \cite{contentCentricRouting}, where both legacy and Wi-Fi interfaces are exploited at group owners (not at clients as in our approach). As compared to this work, our approach (i) uses unicast transmissions (rather than broadcast like \cite{contentCentricRouting}), so our testbed can operate at higher rates, (ii)  supports bidirectional IP communication supporting both TCP and UDP, (iii) requires minimal changes to existing Wi-Fi Direct and legacy Wi-Fi operations. Furthermore, we implement DARS over this testbed by taking into account device capabilities.

%% file: conclusion.tex
\section{\label{sec:conclusion}Conclusion}
In this paper, we developed a {\em device-aware routing and scheduling algorithm} (DARS) over D2D networks by taking into account device capabilities such as computing power, energy, and incentives. Our approach is grounded on a network utility maximization formulation of the problem and its solution. We developed a multi-hop D2D testbed using real mobile devices. We implemented DARS over this testbed. The experimental results demonstrate the benefits of our algorithm.
